\documentclass{cernyrep} 
\usepackage{epsfig}
\usepackage[T1]{fontenc}
\usepackage{graphicx}
\usepackage{epstopdf}
\usepackage{morefloats}
\usepackage{placeins}
                       
\usepackage{varwidth}
\usepackage{xcolor}
\usepackage[customcolors]{hf-tikz}
             
\usepackage{ifthen}
\usepackage{array,tabularx}
\usepackage{amssymb}
\usepackage{xcolor}     
\usepackage{lipsum}     
\usepackage[standard]{ntheorem}   
\usepackage{mdframed}   
\usepackage{enumitem}

\theoremstyle{break}
\theoremheaderfont{\bfseries}
\newmdtheoremenv[%
linecolor=gray,leftmargin=40,%
rightmargin=40,
backgroundcolor=gray!40,%
innertopmargin=10pt,%
ntheorem]{myinterlude}{Interlude}[section]

\newmdtheoremenv[%
linecolor=gray,leftmargin=40,%
rightmargin=40,
backgroundcolor=red!30,%
innertopmargin= 5pt,%
ntheorem]{myinterlude2}{Warning}

\newmdtheoremenv[%
linecolor=gray,leftmargin= 5,%
rightmargin= 5,
backgroundcolor=blue!30,%
innertopmargin= 2pt,%
ntheorem]{myinterlude3}{}

\usepackage[small,bf,nooneline]{caption2}

\newcommand{\Myabstract}[1]{
\begin{quote}
{\large\bfseries{Abstract}}\\
\rule{14cm}{2pt}
\vskip 2mm
#1
\vskip-1mm
\rule{14cm}{2pt}
\end{quote}}

\begin{document}

\section{Imperfections in the Crab-Waist Scheme}     

\noindent
{\it Demin Zhou} \\
{{
KEK, 1-1 Oho, Tsukuba, Ibaraki 305-0801, Japan\\
}}

\Myabstract{
The crab-waist collision scheme has been the baseline choice for SuperKEKB and future circular $e^+e^-$ colliders. Achieved through properly phased sextupoles, the crab-waist transform is essential in suppressing beam-beam resonances, thereby enabling high luminosity in these colliders. In this paper, we explore potential sources of imperfections that may compromise the effectiveness of the crab-waist transform. We begin by reviewing the theoretical framework of the ideal crab-waist scheme and the associated weak-strong beam-beam resonances. Following this, we analyze how machine imperfections could amplify these resonances, thereby impacting collider performance. Finally, we briefly address the connections between theoretical models, simulations, and beam experiments, with a particular focus on the use of weak-strong beam experiments to identify and diagnose potential imperfections in machine settings.
}

\subsection{Introduction}

Since P. Raimondi proposed the crab waist (CW) collision scheme in 2006~\cite{raimondi2006}, and its successful testing at DA$\Phi$NE~\cite{zobov2010test}, extensive investigations over the years have established it as the standard choice for new generation circular $e^+e^-$ colliders. These colliders are designed to achieve luminosities exceeding $10^{35}\ \text{cm}^{-2}\text{s}^{-1}$ for the exploration of new physics beyond the Standard Model.

\begin{figure}[htb]
   \centering
    \vspace{-1mm}
   \includegraphics*[width=0.4\linewidth]{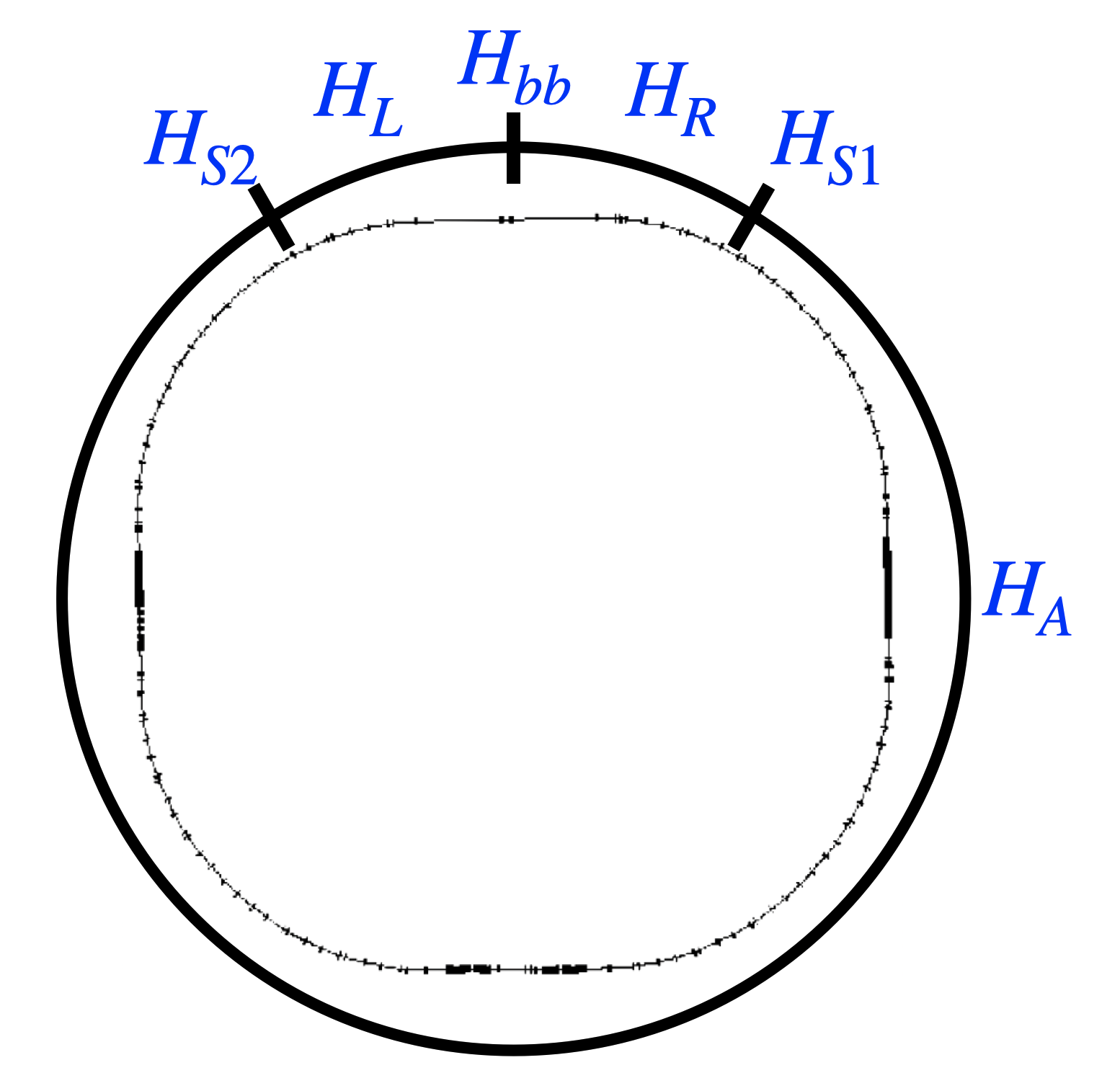}
    \vspace{0mm}
   \caption{A storage ring modeled using Lie maps. The inner ring illustrates the layout of the KEKB LER, while the outer circle represents the sequential maps.}
   \label{LieModel}
\end{figure}

In terms of Lie maps, the one-turn map of a CW collider ring can be expressed:
\begin{equation}
    \mathcal{M}=e^{-:H_R:}e^{-:H_{S1}:}e^{-:H_A:} e^{-:H_{S2}:}e^{-:H_L:}e^{-:H_{bb}:}.
    \label{eq:map1}
\end{equation}
In this context, we assume that the beam starts from the interaction point (IP) and passes sequentially through the following components: the right side of the interaction region (IR, represented by the integrated Hamiltonian $H_R$), the first CW sextupole ($H_{S1}$), the straight and arc sections ($H_A$), the second CW sextupole ($H_{S2}$), the left side of the IR ($H_L$), and finally the beam-beam kick from the opposite beam ($H_{bb}$). The model described by Eq.~\eqref{eq:map1} for a storage ring is illustrated in Fig.~\ref{LieModel}. It is important to note that each map in this formulation can be simplified to include only linear terms or expanded to incorporate specific nonlinear elements, depending on the level of complexity required for the physics being studied.

In the following section, we analyze a simplified version of Eq.~\eqref{eq:map1} for an ideal crab-waist collider, focusing exclusively on the beam dynamics governed by the beam-beam map, $H_{bb}$. We then examine potential imperfections in the crab-waist transform, resulting from irregularities in the other maps within Eq.~\eqref{eq:map1} caused by realistic machine errors. 

\subsection{\label{sec:icwt}Ideal crab-waist transform}

The one-turn map of an ideal CW collider ring is simplified from Eq.~(\ref{eq:map1}) as:
\begin{equation}
    \mathcal{M}_i=e^{-:H_0:}e^{-:H_{cw}:}e^{-:H_{bb}:}e^{:H_{cw}:}.
    \label{eq:map0}
\end{equation}
Here, $H_0$ represents the Hamiltonian for the IP-to-IP lattice transformation. The terms $\pm H_{cw}$ indicate the Hamiltonians for the ideal CW transforms that sandwich the beam-beam interaction at the IP, with $H_{cw}$ expressed as:
\begin{equation}
    H_{cw}=\frac{\chi}{2\tan(2\theta_c)} xp_y^2,
    \label{eq:Hcw}
\end{equation}
where $\chi$ is the relative CW strength acted on the weak beam and $\theta_c$ is the half crossing angle. In the weak-strong model, the weak beam experiences the beam-beam force of the strong beam with its ideal distribution around the IP expressed as:
\begin{equation}
    \rho_0(\Vec{r}_0;s_0)=\frac{e^{-\frac{x_0^2}{2\sigma_{x0}^{*2}}-\frac{y_0^2}{2\sigma_y^2(s_0)}-\frac{z_0^2}{2\sigma_{z0}^2}}}{(2\pi)^{3/2}\sigma_{x0}^*\sigma_y(s_0)\sigma_{z0}},
\end{equation}
where $\Vec{r}_0=(x_0,y_0,z_0)$ indicate the coordinates in the frame of the strong beam, $s_0$ is the orbit length from the IP, and
\begin{equation}
    \sigma_y(s_0)=\sigma_{y0}^*\sqrt{1+\frac{1}{\beta_{y0}^{*2}}\left(z_0+s_0+\frac{\chi_0x_0}{\tan(2\theta_c)} \right)^2}.
\end{equation}
Here, only the spatial dependence of the vertical beam size is significant, since, typically, $\beta_{y0}^* \ll \sigma_{z0} \ll \beta_{x0}^*$. The CW transform tilts the distribution of the strong beam with a relative strength of $\chi_0$, causing the longitudinal waist to be shifted by $s_0=-\chi_0x_0/\tan(2\theta_c)$ with respect to IP. For the full CW transform, $\chi_0=1$.

By rotation transforms, the coordinates in the frame of the strong beam are expressed by the coordinates in the frame of the weak beam. Here, we choose $x_0=-x\cos(2\theta_c)-(z+s)\sin(2\theta_c)$, $z_0+s_0=x\sin(2\theta_c)-(z+s)\cos(2\theta_c)$, $y_0=y$, and $s_0=s$. In the frame of the weak beam, the Hamiltonian for the beam-beam interaction is expressed by $H_{bb}=V_{bb}\delta(s)$, with the beam-beam potential
\begin{align}
    V_{bb} = & -\frac{N_0r_eR_0(1+\cos(2\theta_c))}{\pi\gamma}
    \iiint_{-\infty}^\infty dsdt_xdt_y 
    \frac{\lambda_0(z_0)}{R_0^2t_x^2+t_y^2} \nonumber \\
    & \frac{e^{-\frac{it_xx_0}{\sigma_{x0}^*}-\frac{it_yy_0}{\sigma_{y0}^*}}}{\sqrt{1+\chi_0^2\zeta_{x0}^2t_y^2}}
    e^{-\frac{(t_x+\eta_0t_y^2(z_0+s_0))^2}{2(1+\chi_0^2\zeta_{x0}^2t_y^2)}-\frac{t_y^2}{2}\left( 1+\frac{(z_0+s_0)^2}{\beta_{y0}^{*2}} \right)}.
    \label{eq:Vbb1}
\end{align}
Here, the subscript ``0'' denotes quantities of the strong beam: $N_0$ is the bunch population, $R_0=\sigma_{y0}^*/\sigma_{x0}^*$ is the transverse aspect ratio, $\lambda_0(z_0)$ is the normalized longitudinal Gaussian charge distribution, $\zeta_{x0}=\sigma_{x0}^*/(\beta_{y0}^*\tan(2\theta_c))$, $\eta_0=i\chi_0\zeta_{x0}/\beta_{y0}^*$. $r_e$ is the classical radius of the electron, and $\gamma$ is the relativistic Lorentz factor of the weak beam.

The integral variable $s$ can be replaced by $\tau=-z_0\tan\theta_c/\sigma_{x0}^*$ under the condition that the weak beam particles have negligible variance in their $x$ and $z$ coordinates when experiencing beam-beam kicks around the IP. This implies that $x$ and $z$ do not depend on $s$ in the integral of $V_{bb}$. 
With this replacement, Eq.~(\ref{eq:Vbb1}) can be rewritten as
\begin{align}
    V_{bb} = & -\frac{N_0r_eR_0}{\pi\gamma}
    \iiint_{-\infty}^\infty d\tau dt_xdt_y 
    \frac{\lambda(\tau)}{R_0^2t_x^2+t_y^2}
    e^{it_x(\tau+q_x+\phi_0q_z)} \nonumber \\
    & \frac{e^{-it_yq_y}}{\sqrt{1+\chi_0^2\zeta_{x0}^2t_y^2}}
    e^{-\frac{(t_x-i\chi_0\zeta_{x0}^2t_y^2(\tau+\tau'))^2}{2(1+\chi_0^2\zeta_{x0}^2t_y^2)}-\frac{t_y^2}{2}\left( 1+\zeta_{x0}^2(\tau+\tau')^2 \right)},
    \label{eq:Vbb2}
\end{align}
with $q_x=x/\sigma_{x0}^*$, $q_y=y/\sigma_{y0}^*$, $q_z=z/\sigma_{z0}$, $\tau'=\phi_0q_z-q_x\tan\theta_c\tan(2\theta_c)$, and the Piwinski angle $\phi_0=\sigma_{z0}\tan\theta_c/\sigma_{x0}^*$. The longitudinal charge density, re-normalized with respect to the Piwinski angle, is expressed as
\begin{equation}
    \lambda(\tau)= \frac{1}{\sqrt{2\pi}\phi_0} e^{-\frac{\tau^2}{2\phi_0^2}}.
    \label{eq:Lambdaphi0}
\end{equation}
All quantities within the integral of Eq.~(\ref{eq:Vbb2}) are dimensionless. It can be seen that the geometric parameters $R_0$, $\phi_0$, and $\zeta_{x0}$ fundamentally define the physics of beam-beam effects. In CW colliders, typical conditions are $\theta_c \ll 1$, $R_0 \ll 1$, $\phi_0\gg 1$, and $\zeta_{x0}\lesssim 0.5$. These characteristics form the basis of our discussions in this paper.

We further simplify the problem under discussion by assuming $\chi_0=0$ (that is, the CW transform does not significantly alter the charge distribution of the strong beam, or its impacts on the weak beam are negligible) and approximating $\tau'$ as $\tau' \approx \phi_0q_z$. With these assumptions, Eq.~(\ref{eq:Vbb2}) reduces to
\begin{align}
    V_{bb} = -\frac{N_0r_eR_0}{\pi\gamma}
    \iiint_{-\infty}^\infty d\tau dt_xdt_y 
    \frac{\lambda(\tau)}{R_0^2t_x^2+t_y^2} 
    e^{it_x(\tau+q_x+\phi_0q_z)-it_yq_y}
    e^{-\frac{t_x^2}{2}-\frac{t_y^2}{2}\left( 1+\zeta_{x0}^2(\tau+\phi_0q_z)^2 \right)}.
    \label{eq:Vbb3}
\end{align}

The motions of the weak beam particles around the IP can be described by
\begin{equation}
    x=\sqrt{2\beta_x^*J_x}\cos\psi_x,
    \label{eq:xmotion}
\end{equation}
\begin{equation}
    y(s')=\sqrt{2\beta_y(s')J_y}\cos\phi_y(s'),
    \label{eq:ymotion-hr}
\end{equation}
\begin{equation}
    z=\sqrt{2\beta_zJ_z}\cos\psi_z,
    \label{eq:zmotion}
\end{equation}
with
\begin{equation}
    \beta_y(s')=\beta_y^*\left( 1+\frac{1}{\beta_y^{*2}} \left( s'+ \frac{\chi x}{\tan(2\theta_c)} \right)^2 \right),
    \label{eq:betay-hr}
\end{equation}
\begin{equation}
    \phi_y(s')=\psi_y+\arctan\left( \frac{s'+\frac{\chi x}{\tan(2\theta_c)}}{\beta_y^*} \right),
\end{equation}
with $s'=z+s$. For simplification, it is assumed that the CW transform and the hourglass effects are considered only in the vertical direction (this is the same as the strong beam). Using the definition of $\tau$, we have
\begin{equation}
    s'=\frac{1}{\sin(2\theta_c)}\left[ \sigma_{x0}^*\tau + z\tan\theta_c + x\tan\theta_c\sin(2\theta_c) \right].
\end{equation}
The term involving $x$ is small compared to the other two terms. Therefore, in the following analysis, we use the approximation $s'\approx (\sigma_{x0}^*\tau + z\tan\theta_c)/\sin(2\theta_c) $.

Using the canonical transform, the beam-beam Hamiltonian is rewritten as $H_{bb}=V_{bb}\delta(\theta)$ with $\theta=2\pi s/C$ where $C$ is the circumference of the ring. The beam-beam Hamiltonian is further expanded into a sum of Fourier modes:
\begin{equation}
    V_{bb}\delta(\theta)=\sum_{\vec{m},n}
    V_{m_xm_ym_z}e^{i(m_x\psi_x+m_y\psi_y+m_z\psi_z-n\theta)},
    \label{eq:Vbb4}
\end{equation}
with $\vec{m}=(m_x,m_y,m_z)$ and the amplitude term calculated by
\begin{align}
    V_{m_xm_ym_z}=& \frac{1}{(2\pi)^4}
    \iiint_0^{2\pi} d\psi_xd\psi_yd\psi_z
    V_{bb} e^{-i(m_x\psi_x+m_y\psi_y+m_z\psi_z)}.
    \label{eq:Vmxmymz1}
\end{align}
In general, the amplitude terms $V_{m_xm_ym_z}$ characterize various aspects of the beam-beam effects: amplitude-dependent tune shifts (i.e., $V_{000}$), 1D and 2D betatron resonances (i.e., $V_{m_x00}$, $V_{0m_y0}$, and $V_{m_xm_y0}$), 2D and 3D synchrobetatron resonances (i.e., $V_{m_x0m_z}$, $V_{0m_ym_z}$, and $V_{m_xm_ym_z}$), etc..

\begin{figure}[htb]
   \centering
    \vspace{-1mm}
   \includegraphics*[width=0.7\linewidth]{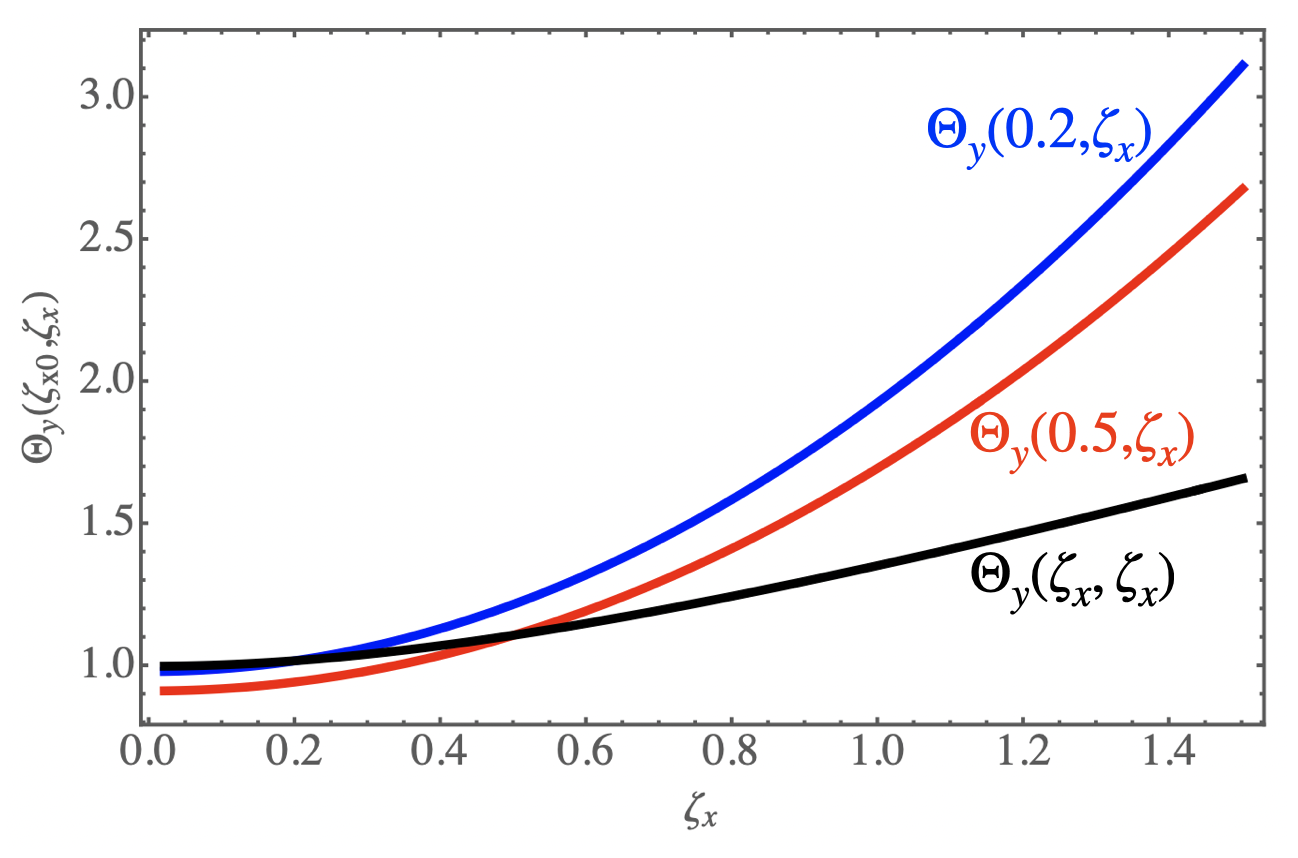}
    \vspace{0mm}
   \caption{Variation of the vertical hourglass factor as a function of $(\zeta_{x0}, \zeta_x)$, as described in Eq.~(\ref{eq:thetay}).}
   \label{ThetaY}
\end{figure}

The beam-beam strength parameters can be calculated from $V_{000}$ with Eq.~(\ref{eq:Vbb3}) as
\begin{equation}
    \xi_x=\frac{N_0r_e\beta_x^*}{2\pi\gamma\overline{\sigma}_{x0}(\overline{\sigma}_{x0}+\sigma_{y0}^*)},
\end{equation}
\begin{equation}
    \xi_y=\frac{N_0r_e\beta_y^*}{2\pi\gamma\sigma_{y0}^*(\overline{\sigma}_{x0}+\sigma_{y0}^*)} \Theta_y(\zeta_{x0},\zeta_x),
    \label{eq:xiy}
\end{equation}
\begin{equation}
    \xi_z=\frac{N_0r_e\beta_z\tan^2\theta_c}{2\pi\gamma\overline{\sigma}_{x0}(\overline{\sigma}_{x0}+\sigma_{y0}^*)},
\end{equation}
with the hourglass factor in the vertical direction given by
\begin{equation}
    \Theta_y=
    \frac{e^{u_0}}{2\sqrt{2\pi}\zeta_{x0}^3}
    \left[ (2\zeta_{x0}^2-\zeta_x^2)K_0\left( u_0 \right)
    + \zeta_x^2 K_1\left( u_0 \right) \right],
    \label{eq:thetay}
\end{equation}
where $\overline{\sigma}_{x0}=\sqrt{\sigma_{x0}^{*2}+\sigma_{z0}^2\tan^2\theta_c}$, $u_0=1/(4\zeta_{x0}^2)$, and $\zeta_x=\sigma_{x0}^*/(\beta_y^*\tan(2\theta_c))$. $K_0(x)$ and $K_1(x)$ are modified Bessel functions of the second kind. The hourglass effects on the horizontal and longitudinal beam-beam parameters are largely negligible in CW colliders. As shown in Fig.~\ref{ThetaY}, $\Theta_y(\zeta_{x0},\zeta_x)$ remains close to 1 when $\zeta_{x0}=\zeta_x<0.5$; however, for larger values, the hourglass effect can significantly increase the vertical beam-beam parameter. This indicates that $\beta_{y0}^*=\beta_y^*>\sigma_{x0}^*/\theta_c$ should be considered as the hourglass condition for CW colliders.

\begin{figure}[htb]
   \centering
    \vspace{-1mm}
   \includegraphics*[width=0.7\linewidth]{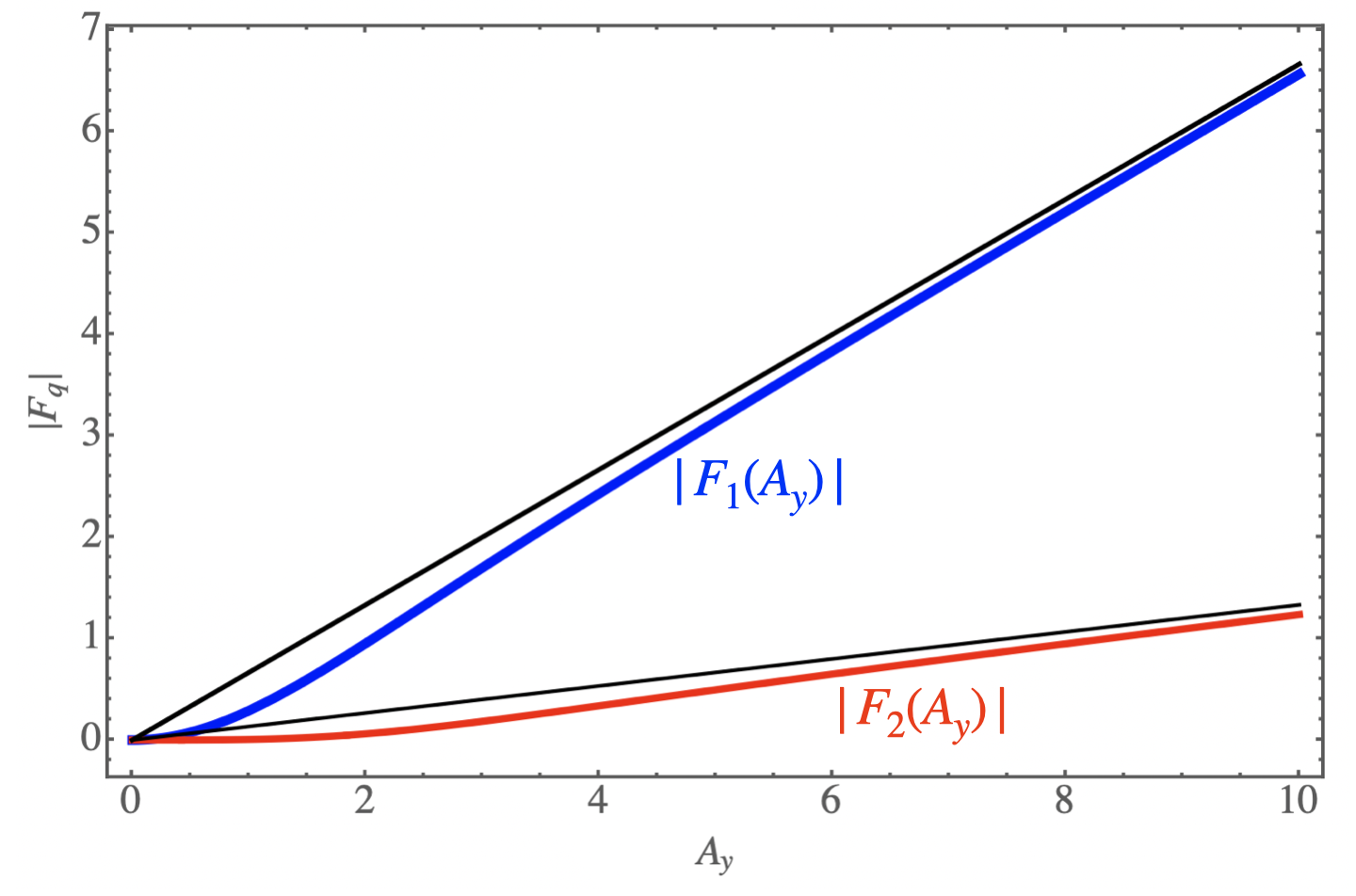}
    \vspace{0mm}
   \caption{Dependence of $|F_q|$ on the vertical normalized amplitude. Blue and red lines correspond to $q=1$ and $q=2$, respectively. The black lines show the large-$A_y$ approximations as given by Eq.~(\ref{eq:Fq-large-Ay}).}
   \label{Fq}
\end{figure}

We follow the formalism of~\cite{dikansky2009} to investigate the resonances of $V_{m_xm_ym_z}$ for cases where $m_y=2q>0$, as resonances with odd $m_y$ vanish under the formulation given by Eq.~(\ref{eq:Vbb3}). In such cases, the term $1/(R_0^2t_x^2+t_y^2)$ can be approximated as $1/t_y^2$. By changing the integral variables from $\tau$ and $t_y$ to $r=\tau+\phi_0q_z+q_x$ and $u=t_y\sqrt{1+\zeta_x^2(\tau+\phi_0q_z+\chi q_x)^2}$, we derive the explicit expression for $V_{m_xm_ym_z}$:
\begin{align}
    V_{m_xm_ym_z}= & -\frac{N_0r_eR_0}{\pi \gamma (2\pi)^3} \iint_0^{2\pi} d\psi_x d\psi_z e^{-i(m_x\psi_x+m_z\psi_z)} 
    \sqrt{2\pi} \int_{-\infty}^\infty dr \sqrt{1+\zeta_x^2(r+(\chi-1)q_x)^2} \nonumber \\
    & \lambda(r-q_x-\phi_0q_z) e^{-\frac{r^2}{2}} 
    \left( \frac{1+i\zeta_x(r+(\chi-1)q_x)}{1-i\zeta_x(r+(\chi-1)q_x)} \right)^q
    F_q(A_y),
    \label{eq:Vmxmymz2}
\end{align}
with
\begin{align}
    F_q(A_y) = (-1)^q \int_{-\infty}^\infty du \frac{1}{u^2} J_{2q}(A_yu) e^{-\frac{\Lambda u^2}{2}},
\end{align}
where $A_y=\sqrt{2\beta_y^*J_y}/\sigma_{y0}^*$ is the normalized vertical amplitude and 
\begin{equation}
    \Lambda(x)=\frac{1+\zeta_{x0}^2(r-q_x)^2}{1+\zeta_x^2(r+(\chi-1)q_x)^2}.
\end{equation}
Here, $J_{n}(x)$ is the Bessel function of the first kind. Following~\cite{dikansky2009}, we approximate $\Lambda(x)\approx 1$ (this is valid when $\beta_{y}^*=\beta_{y0}^*$ and the CW modulation on the strong beam is considered), leading to the following expression:
\begin{align}
    F_q(A_y)\approx 
    \frac{(-1)^q}{4q^2-1} \sqrt{\frac{\pi}{2}} e^{-\frac{A_y^2}{4}}
    \left[ (2+4q+A_y^2)I_q\left(\frac{A_y^2}{4}\right) +A_y^2 I_{q+1}\left(\frac{A_y^2}{4}\right) \right],
    \label{eq:Fq}
\end{align}
where $I_q(x)$ represents the modified Bessel function of the first kind. For large and small values of $A_y$, $F_q(A_y)$ can be approximated as follows:
\begin{equation}
    F_q(A_y)\approx (-1)^q \frac{2A_y}{4q^2-1} \ \text{for} \ A_y\gg 1,
    \label{eq:Fq-large-Ay}
\end{equation}
\begin{equation}
    F_q(A_y)\approx \frac{(-1)^q2^{1-3q}}{q!(2q-1)} \sqrt{\frac{\pi}{2}} A_y^{2q} e^{-\frac{A_y^2}{4}} \ \text{for} \ A_y \ll 1.
\end{equation}
For $q=m_y/2=1,2$, the dependence of $|F_q|$ on the amplitude of the vertical oscillations is shown in Fig.~\ref{Fq}.

The dependence of the longitudinal coordinate $z$ in Eq.~(\ref{eq:Vmxmymz2}) only appears in $\lambda(r-q_x-\phi_0q_z)$. Then, utilizing the identity of 
\begin{equation}
    \lambda(\tau)=\frac{1}{2\pi\phi_0} \int_{-\infty}^\infty dk e^{\frac{ik\tau}{\phi_0} - \frac{k^2}{2}},
\end{equation}
the integral over $\psi_z$ can be performed, yielding
\begin{equation}
    V_{m_xm_ym_z} = -\frac{N_0r_eR_0}{\pi\gamma\phi_0}
    F_q(A_y) \overline{G}_{m_xm_ym_z}(A_x,A_z),
    \label{eq:Vmxmymz3}
\end{equation}
with
\begin{equation}
    \overline{G}_{m_xm_ym_z}(A_x,A_z) = 
    \int_{-\infty}^\infty \frac{d\psi_x}{2\pi} e^{-im_x\psi_x}
    G_{m_ym_z}(q_x,A_z),
\end{equation}
\begin{align}
    G_{m_ym_z}(q_x,A_z) = 
    & \int_{-\infty}^\infty \frac{dr}{\sqrt{2\pi}} e^{-\frac{r^2}{2}} \mathscr{I}_{m_z}(A_z,r-q_x) \nonumber \\
    & \sqrt{1+\zeta_x^2(r+(\chi-1)q_x)^2} 
    \left( \frac{1+i\zeta_x(r+(\chi-1)q_x)}{1-i\zeta_x(r+(\chi-1)q_x)} \right)^q,
\end{align}
\begin{equation}
    \mathscr{I}_{m_z}(A_z,r-q_x) = 
    \frac{(-i)^{m_z}}{2\pi}
    \int_{-\infty}^\infty dk J_{m_z}(kA_z) e^{\frac{ik(r-q_x)}{\phi_0}-\frac{k^2}{2}},
\end{equation}
where $A_z=\sqrt{2\beta_zJ_z}/\sigma_{z0}$ is the normalized longitudinal amplitude.

For $\chi=0$ (i.e., without CW) and $\phi_0 \gg 1$, $\mathscr{I}_{m_z}(A_z,r-q_x)$ can be approximated by $\mathscr{I}_{m_z}(A_z,0)$, given by:
\begin{equation}
    \mathscr{I}_{m_z}(A_z,0) = \frac{(-i)^{m_z} \left[ 1+(-1)^{m_z} \right]}{2\sqrt{2\pi}}
    e^{-\frac{A_z^2}{4}} I_{\frac{m_z}{2}}\left( \frac{A_z^2}{4} \right).
\end{equation}
Consequently, we obtain
\begin{equation}
    G_{m_ym_z}(q_x,A_z) \approx \mathscr{I}_{m_z}(A_z,0) Y_q(q_x,\zeta_x),
\end{equation}
where
\begin{align}
    Y_q(q_x,\zeta_x) = 
    \int_{-\infty}^\infty \frac{dr}{\sqrt{2\pi}} e^{-\frac{r^2}{2}}
    \sqrt{1+\zeta_x^2(r-q_x)^2}
    \left( \frac{1+i\zeta_x(r-q_x)}{1-i\zeta_x(r-q_x)} \right)^q,
\end{align}
and
\begin{equation}
    \overline{G}_{m_xm_ym_z}(A_x,A_z) \approx \mathscr{I}_{m_z}(A_z,0) \overline{Y}_{qm_x}(A_x,\zeta_x),
    \label{eq:GxyznoCW}
\end{equation}
with
\begin{align}
    \overline{Y}_{qm_x}(A_x,\zeta_x) =
    & \int_{-\infty}^\infty \frac{dr}{\sqrt{2\pi}} e^{-\frac{r^2}{2}} \int_{-\infty}^\infty \frac{d\psi_x}{2\pi} e^{-im_x\psi_x} \nonumber \\
    & \sqrt{1+\zeta_x^2(r-A_x \cos\psi_x)^2} 
    \left( \frac{1+i\zeta_x(r-A_x \cos\psi_x)}{1-i\zeta_x(r-A_x \cos\psi_x)} \right)^q,
\end{align}
where $A_x=\sqrt{2\beta_x^*J_x}/\sigma_{x0}^*$ is the normalized horizontal amplitude. Equation~(\ref{eq:GxyznoCW}) indicates that, without CW, non-vanishing beam-beam resonances occur when $m_x=\textit{integer}$, $m_y=\textit{even}$, and $m_z=\textit{even}$.

For $\chi=1$ (i.e., full CW) and $\phi_0 \gg 1$, $\mathscr{I}_{m_z}(A_z,r-q_x)$ can be approximated by $\mathscr{I}_{m_z}(A_z,-q_x)$. Consequently, we obtain
\begin{equation}
    \overline{G}_{m_xm_ym_z}^{cw}(A_x,A_z) \approx \overline{\mathscr{I}}_{m_xm_z}(A_x,A_z) Y_{q}(0,\zeta_x),
    \label{eq:GxyzCW}
\end{equation}
where
\begin{align}
    \overline{\mathscr{I}}_{m_xm_z}(A_x,A_z) =
    (-i)^{m_x+m_z} \int_{-\infty}^\infty \frac{dk}{2\pi} e^{-\frac{k^2}{2}}
    J_{m_x}\left( \frac{kA_x}{\phi_0} \right) J_{m_z}(kA_z).
\end{align}
Equation~(\ref{eq:GxyzCW}) indicates that, with full CW, non-vanishing beam-beam resonances occur when $m_x+m_z=\textit{even}$ and $m_y=\textit{even}$.

According to Eq.~(\ref{eq:Vmxmymz3}), the dependence of $V_{m_xm_ym_z}$ on $A_y$ remains the same with and without CW. In the following subsections, we will analyze the term of $\overline{G}_{m_xm_ym_z}(A_x,A_z)$ to separately discuss 2D betatron resonances $V_{m_xm_y0}$, 2D horizontal synchrobetatron resonances $V_{m_x0m_z}$, 2D vertical synchrobetatron resonances $V_{0m_ym_z}$, and 3D synchrobetatron resonances $V_{m_xm_ym_z}$ in CW colliders.

\subsubsection{Betatron resonances $V_{m_xm_y0}$}

\begin{figure}[htb]
   \centering
    \vspace{-1mm}
   \includegraphics*[width=0.7\linewidth]{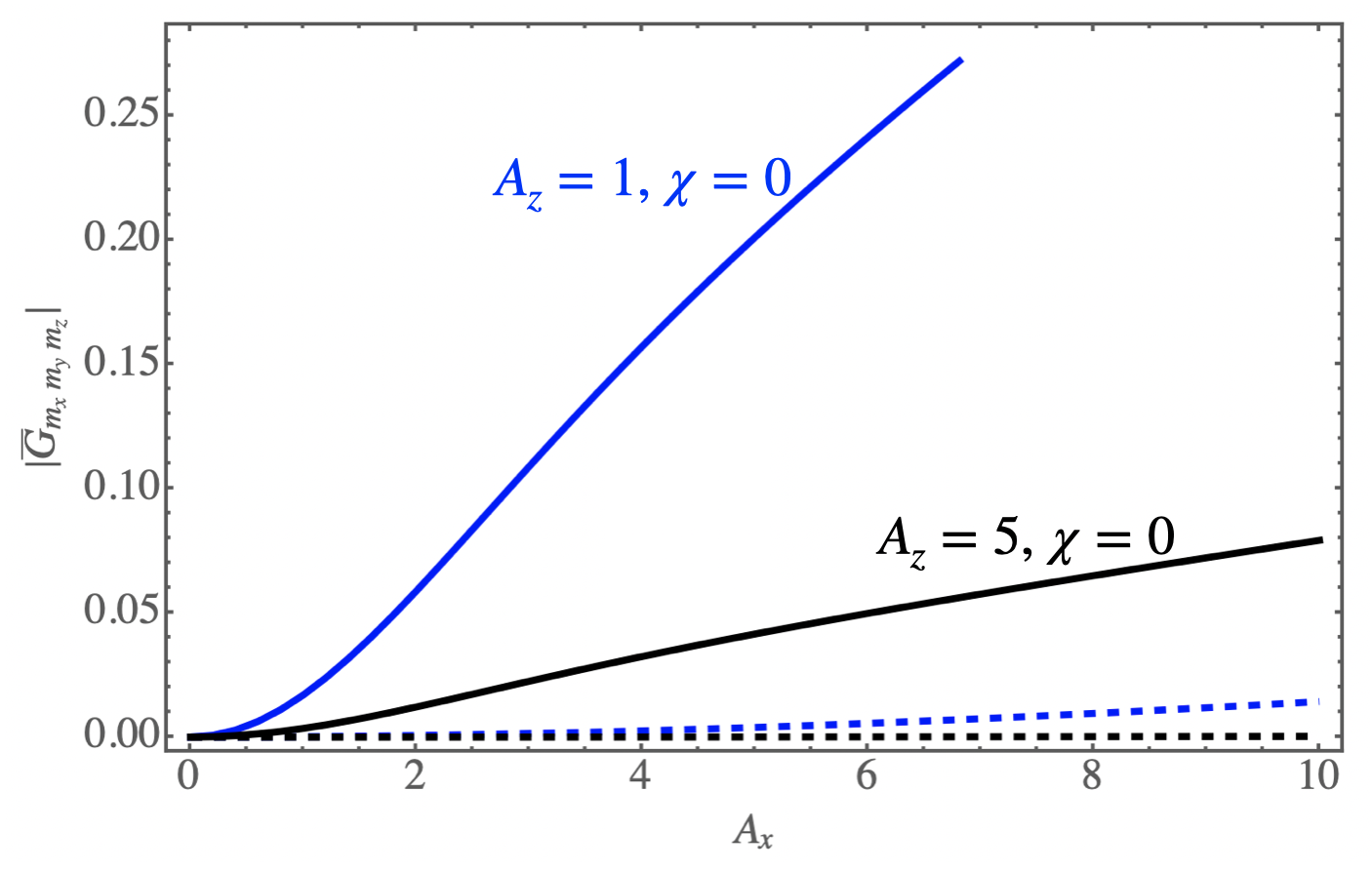}
    \vspace{0mm}
   \caption{Dependence of $|\overline{G}_{220}|$ and $|\overline{G}_{220}^{cw}|$ on the horizontal and longitudinal normalized amplitudes with $\zeta_x=0.5$ and $\phi_0=10$. The solid blue and black lines correspond to $A_z=1$ and $A_z=5$, respectively, without CW. The dashed blue and black lines correspond to $A_z=1$ and $A_z=5$, respectively, with CW.}
   \label{Ga220}
\end{figure}

Betatron resonances with $m_x=1,2,3,\ldots$ and $m_y=2,4,6,\ldots$ can be excited by beam-beam interaction without CW. However, with CW, the resonances with $m_x=1,3,5,\ldots$ will not be excited. Those with $m_x=2,4,6,\ldots$ will be significantly suppressed, although they still have a finite amplitude that depends on the amplitude of longitudinal oscillations. This suppression effect is demonstrated in Fig.~\ref{Ga220} using the parameters $\zeta_x=0.5$ and $\phi_0=10$ for the remaining lowest-order resonance with $(m_x,m_y,m_z)=(2,2,0)$.

\subsubsection{Horizontal synchrobetatron resonances $V_{m_x0m_z}$}

The resonances of $V_{m_x0m_z}$ and their impact on horizontal emittance growth are analyzed in~\cite{kicsiny2024}. It is demonstrated that Eq.~(\ref{eq:Vmxmymz2}) does not hold for small $A_y$ and $m_y=0$. In such cases, the term $1/(R_0^2t_x^2+t_y^2)$ must be retained. Taking the limit as $A_y\rightarrow 0$ and $m_y=0$, we derive the following from Eq.~(\ref{eq:Vmxmymz1}) for $m_x+m_z=\textit{even}$:
\begin{equation}
    V_{m_x0m_z} \approx
    -\frac{N_0r_e}{\pi\gamma} i^{m_x+m_z} F_{m_xm_z} (A_x,A_z),
\end{equation}
where
\begin{align}
    F_{m_xm_z} (A_x,A_z) =
    \int_0^\infty \frac{dk}{k} e^{-\frac{k^2}{2}}
    J_{m_x} \left( \frac{kA_x}{\sqrt{\phi_0^2+1}} \right)
    J_{m_z} \left( \frac{k\phi_0A_z}{\sqrt{\phi_0^2+1}} \right).
    \label{eq:Fmxmz1}
\end{align}
For $m_x+m_z=\textit{odd}$, $V_{m_x0m_z}=0$. To arrive at the above equation, the condition $R_0 \ll 1$ is applied, which renders the amplitude $V_{m_x0m_z}$ insensitive to $R_0$, $\zeta_{x0}$, and $\chi$. This indicates that CW does not suppress horizontal synchrobetatron resonances. The strength of these resonances depends solely on the Piwinski angle $\phi_0$ and is not affected by $R_0$ and $\beta_y^*$.

\begin{figure}[htb]
   \centering
    \vspace{-1mm}
   \includegraphics*[width=0.7\linewidth]{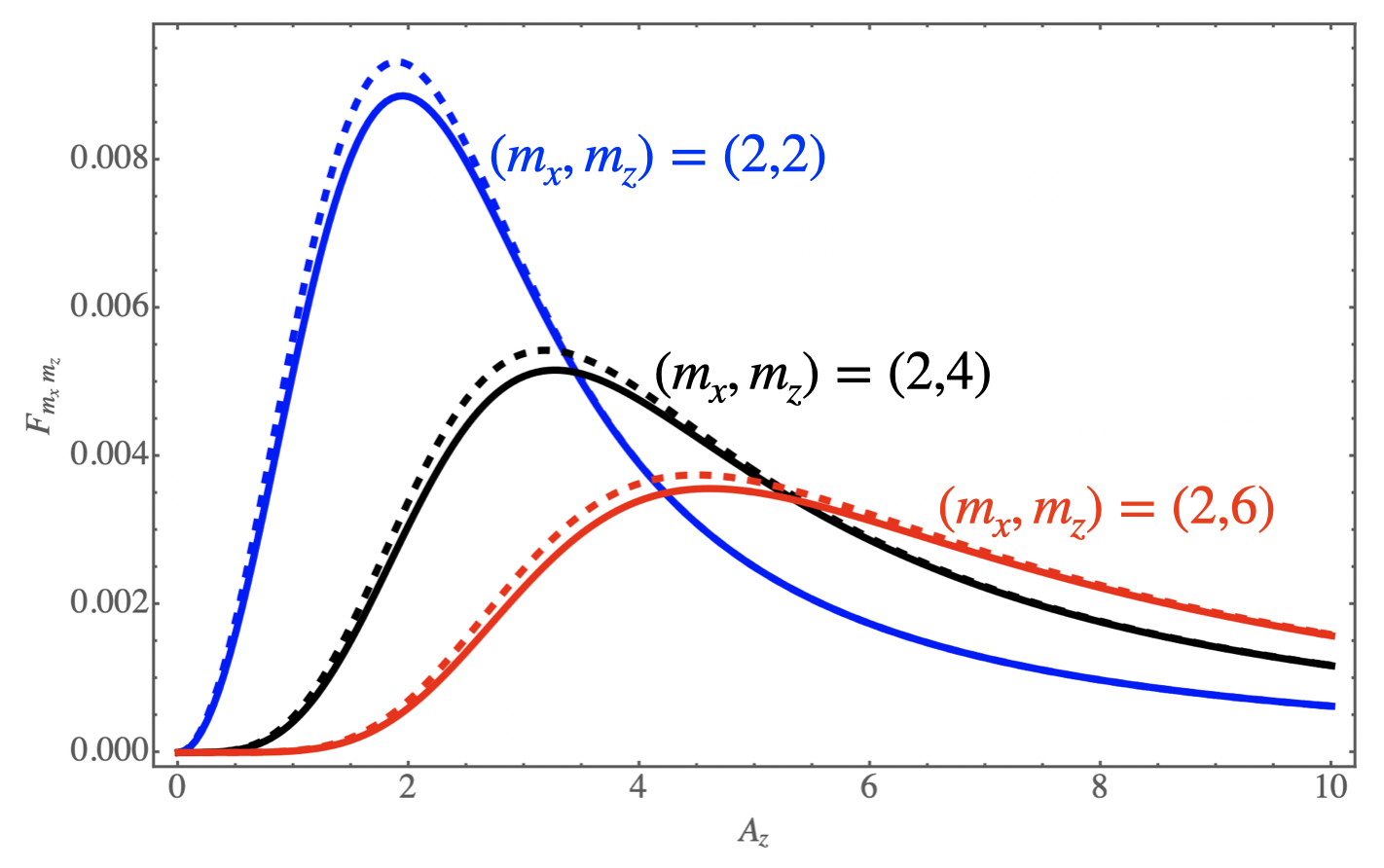}
    \vspace{0mm}
   \caption{Dependence of $F_{2m_z}$ on the longitudinal normalized amplitudes with $A_x=5$ and $\phi_0=10$. The solid and dashed lines correspond to Eqs.~(\ref{eq:Fmxmz1}) and~(\ref{eq:F2mz1}), respectively.}
   \label{F2mz}
\end{figure}

The working points of the CW $e^+e^-$ colliders are typically close to half-integer values in the tune space of $(\nu_x,\nu_y)$. As a result, the resonances $V_{m_x0m_z}$ with $m_x=2$ and $m_z=2,4,6,\ldots$ become particularly significant. Consider $\phi_0\gg 1$ and $m_x=2$, $F_{m_xm_z} (A_x,A_z)$ can be approximated by
\begin{align}
    F_{2m_z}^a \approx 
    \frac{\sqrt{2\pi}}{32\phi_0^2} A_x^2 A_z
    e^{-\frac{A_z^2}{4}} \left[ I_{\frac{m_z-1}{2}} \left( \frac{A_z^2}{4} \right) - I_{\frac{m_z+1}{2}} \left( \frac{A_z^2}{4} \right) \right].
    \label{eq:F2mz1}
\end{align}
Consider $A_x=5$ and $\phi_0=10$. Equations~(\ref{eq:Fmxmz1}) and~(\ref{eq:F2mz1}) are compared in Fig.~\ref{F2mz}, demonstrating general good agreement. For a large Piwinski angle $\phi_0$, Eq.~(\ref{eq:F2mz1}) suggests that the strength of horizontal synchrobetatron resonances is directly proportional to $(A_x/\phi_0)^2$, and that the maximum amplitude shifts to higher $A_z$ values as the index $m_z$ increases. These findings qualitatively align with the simulation results presented in~\cite{kicsiny2024}.


\subsubsection{Vertical synchrobetatron resonances $V_{0m_ym_z}$}

\begin{figure}[htb]
   \centering
    \vspace{-1mm}
   \includegraphics*[width=0.7\linewidth]{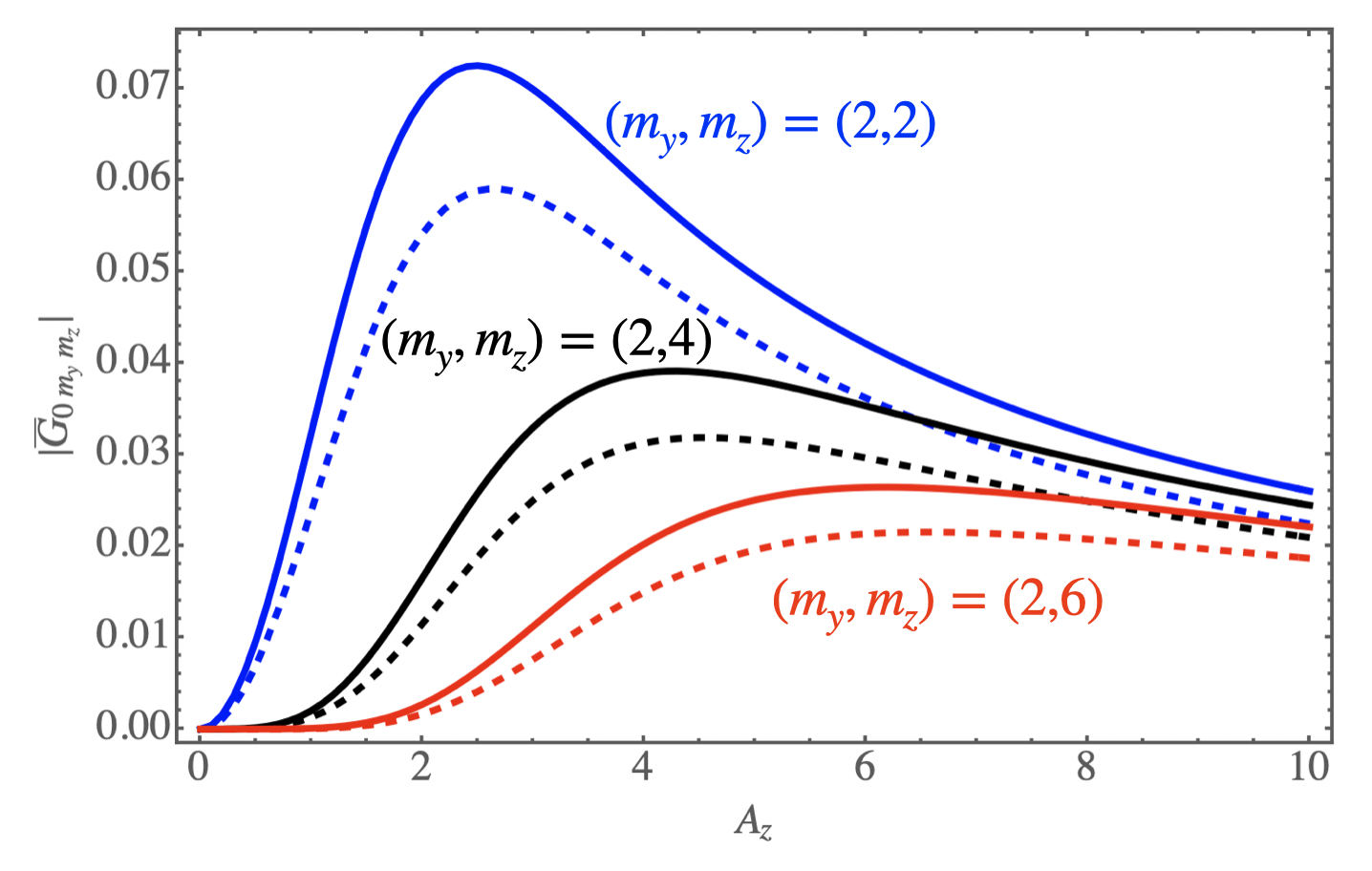}
    \vspace{0mm}
   \caption{Dependence of $|\overline{G}_{02m_z}|$ on the longitudinal normalized amplitudes with $A_x=5$, $\zeta_x=0.5$, and $\phi_0=10$. The solid and dashed lines correspond to without CW ($\chi=0$) and full CW ($\chi=1$), respectively.}
   \label{Ga02mz}
\end{figure}

Vertical synchrobetatron resonances with $m_y=2,4,6,\ldots$ and $m_z=2,4,6,\ldots$ can be excited. Although CW has some suppressive effect on particles with large horizontal amplitudes, it is not effective in fully suppressing these resonances. This behavior is illustrated in Fig.~\ref{Ga02mz} for the case of $m_y=2$.

\subsubsection{3D synchrobetatron resonances $V_{m_xm_ym_z}$}

\begin{figure}[htb]
   \centering
    \vspace{-1mm}
   \includegraphics*[width=0.7\linewidth]{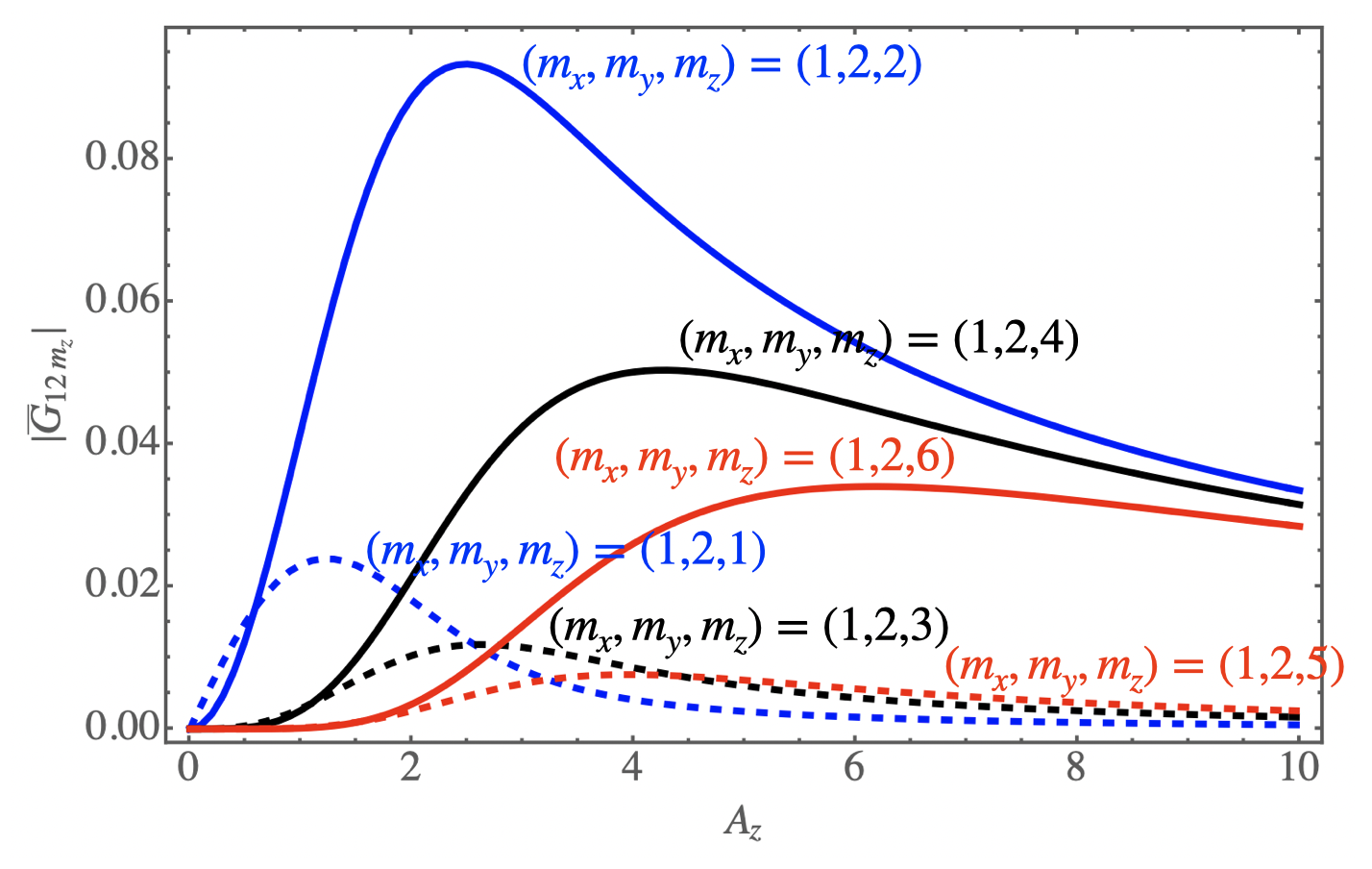}
    \vspace{0mm}
   \caption{Dependence of $|\overline{G}_{12m_z}|$ on the longitudinal normalized amplitudes with $A_x=5$, $\zeta_x=0.5$, and $\phi_0=10$. The solid lines correspond to without CW ($\chi=0$) and $m_z=2,4,6$. The dashed lines correspond to with full CW ($\chi=1$) and $m_z=1,35$.}
   \label{Ga12mz}
\end{figure}

Betatron resonance satellites of the form $m_x\nu_x+m_y\nu_y+m_z\nu_z=\textit{integer}$ can be excited. In the absence of CW, only satellites with $m_z=\textit{even}$ can be excited. However, with CW, only satellites where $m_x+m_z=\textit{even}$ are excited. This means that when $m_x=\textit{odd}$, only satellites with odd $m_z=\textit{odd}$ are excited. This suggests that CW can excite certain 3D synchrobetatron resonances, which may be particularly harmful if the lattice has chromatic aberrations. These characteristics are illustrated in the case of $(m_x,m_y)=(1,2)$, as shown in Fig.~\ref{Ga12mz}.

\subsection{\label{sec:mi}Machine imperfections}

The beam-beam resonances discussed in the previous section can influence beam dynamics, potentially requiring careful machine tunings to maintain optimal luminosity in CW colliders. In these colliders, the impact of imperfections is amplified by the extremely small $\beta_y^*$ and vertical emittance. Beam-beam resonances can be viewed as an intrinsic form of imperfection within the CW transform for CW colliders. Conversely, the linear or nonlinear maps in Eq.~\eqref{eq:map1}, which cause $\mathcal{M}$ to differ from $\mathcal{M}_i$, can be considered as machine imperfections if they amplify the strengths of beam-beam resonances, and as machine optimizations if they further suppress these resonances. In the following, we outline some machine imperfections that should be addressed during the design or operational phases of CW colliders, while reserving the discussion of machine optimizations for another paper.

\subsubsection{Realistic crab-waist transform using sextupoles}

Consider a dedicated thin-lens normal sextupole for $H_{S1}$ in Fig.~\ref{LieModel} to implement the crab-waist transformation. The expression for $H_{S1}$ is given by
\begin{equation}
    H_{S1} = \frac{1}{6}K_2 \left( x^3-3xy^2 \right),
    \label{eq:Hsext1}
\end{equation}
where $K_2$ represents the integrated strength of the magnet. With the horizontal and vertical phase advances between the IP and the sextupole set to $\pi$ and $\pi/2$, respectively (see Sec. 2.5.2, ``\textit{Crab Waist Collision Scheme}'' in~\cite{handbook2023} for the ideal configuration of the crab-waist scheme), the effective Hamiltonian for $H_{S1}$, translated to the IP, is given by
\begin{equation}
    H_{cw}^e=\frac{1}{2}K_2 \beta_y \beta_y^* \sqrt{\frac{\beta_x}{\beta_x^*}} xp_y^2 -\frac{1}{6} K_2 \sqrt{\frac{\beta_x^3}{\beta_x^{*3}}} x^3,
    \label{eq:Hcwe1}
\end{equation}
where $\beta_{x,y}$ represent the $\beta$-functions at the sextupole. The first term on the right-hand side of Eq.~\eqref{eq:Hcwe1} is intended to achieve Eq.~\eqref{eq:Hcw}, under the condition that
\begin{equation}
    K_2=\frac{1}{\tan(2\theta_c)} \frac{1}{\beta_y\beta_y^*} \sqrt{\frac{\beta_x^*}{\beta_x}}.
    \label{eq:k2cw}
\end{equation}
However, the second term on the right-hand side of Eq.~\eqref{eq:Hcwe1} is undesirable as it modifies the particle motion described by Eqs.~\eqref{eq:xmotion}-\eqref{eq:zmotion} and may enhance horizontal synchrobetatron resonances induced by beam-beam interactions.

\subsubsection{\label{sec:ipi}Orbit distortion and linear aberrations at the IP}

The closed orbit offsets and linear optics aberrations at the IP for the weak beam alter the equations of motion, as shown in Eqs.~\eqref{eq:xmotion}-\eqref{eq:zmotion}. A vertical orbit offset leads to luminosity loss, described by
\begin{equation}
    L=L_0 e^{-\frac{\Delta y^2}{2\left(\sigma_{y0}^{*2}+\sigma_y^{*2}\right)}},
\end{equation}
which indicates that the luminosity of CW colliders is highly sensitive to vertical orbit offsets at the IP, primarily due to the extremely small beam sizes. Furthermore, it also results in the excitation of beam-beam resonances with $m_y=\textit{odd}$, according to Eq.~\eqref{eq:Vbb3}.

Horizontal or longitudinal offset excites horizontal synchrobetatron resonances with $m_x+m_z=\textit{odd}$. In this scenario, Eq.~\eqref{eq:Fmxmz1} is modified to
\begin{align}
    F_{m_xm_z} (A_x,A_z) = \frac{1}{2}
    \int_{-\infty}^\infty \frac{dk}{|k|} e^{-\frac{k^2}{2}+\frac{ik}{\sqrt{\phi_0^2+1}}\left( \frac{\Delta x}{\sigma_{x0}^*} + \frac{\phi_0 \Delta z}{\sigma_{z0}} \right)}
    J_{m_x} \left( \frac{kA_x}{\sqrt{\phi_0^2+1}} \right)
    J_{m_z} \left( \frac{k\phi_0A_z}{\sqrt{\phi_0^2+1}} \right).
    \label{eq:Fmxmz2}
\end{align}
In particular, the potential-well distortion caused by longitudinal impedance can lead to a longitudinal shift in bunch center as bunch currents increase. Horizontal beam-size blowup, due to horizontal synchrobetatron resonances $m_x+m_z=\textit{odd}$ was observed in SuperKEKB and studied in detail by simulations and theoretical analysis in~\cite{zhou2023prab, kicsiny2024}. Equation~\eqref{eq:Fmxmz2} suggests two methods to mitigate the longitudinal offset caused by potential-well distortion: one is by adjusting the RF phase and the other by shifting the horizontal offset at IP. The latter requires a relatively large horizontal offset, as given by $\Delta x/\sigma_{x0}^* = -\phi_0 \Delta z/\sigma_{z0}$, since $\phi_0 \gg 1$ and $\Delta z/\sigma_{z0}$ can be on the order of 1 (see, for example,~\cite{zhou2024nima}).

The ideal IP-to-IP lattice transform, represented by $H_0$ in Eq.~\eqref{eq:map0}, should only contain information about the $\beta$ functions at the IP and the one-turn phase advances. Any other linear optics functions, such as $\alpha$ functions, linear betatron couplings, dispersion functions, and other longitudinal-to-transverse couplings, should be treated as linear aberrations at the IP. These aberrations modify Eqs.~\eqref{eq:xmotion}-\eqref{eq:zmotion}, introducing additional resonance terms (see Sec. 2.5.1, ``\textit{Beam-Beam Interactions in Circular Colliders}'' in~\cite{handbook2023}).

Closed orbit distortion and linear aberrations at the IP have been recognized as crucial factors in achieving optimal luminosity in KEKB~\cite{tawada2005mac}, and play an even more significant role in SuperKEKB~\cite{zhou2024jst}.

\subsubsection{Dynamic beta and dynamic emittance due to the beam-beam force}

In a weak-strong model, the beam-beam force can be treated as a standard magnetic element within the ring. In this case, the linear beam-beam force perturbs the linear optics, leading to what are known as dynamic beta and dynamic emittance to the beam. From the perspective of beam-beam resonances, this linear beam-beam effect alters the beam distribution near the IP, as described in Eq.~\eqref{eq:Vbb1}, and consequently influences the beam-beam resonances, as formulated in Sec.~\ref{sec:icwt}.

\subsubsection{\label{sec:cwsexti}Orbit distortion and linear aberrations at the crab-waist sextupoles}

In addition to optics aberrations at the IP, closed-orbit distortion and linear optics aberrations at the crab-waist sextupoles are also critical factors affecting the effectiveness of the crab-waist transformation. The closed-orbit distortion modifies the Hamiltonian of $H_{S1}$ to
\begin{equation}
    H_{S1} = \frac{1}{6}K_2 \left[ (x+\Delta x)^3-3(x+\Delta x)(y+\Delta y)^2 \right],
    \label{eq:Hsext2}
\end{equation}
resulting in the so-called ``feed-down'' effects on linear optics: the horizontal orbit offset induces beta-beat, while the vertical orbit offset generates linear X-Y coupling. It was found that a horizontal closed-orbit offset on the order of 10 $\mu$m at the crab-waist sextupoles in SuperKEKB, with $\beta_y^*=1$ mm, can cause a significant tune shift and beta-beat in the vertical plane~\cite{ohnishi2022mac, sugimoto2022mac}.

The beta-beat at the crab-waist sextupoles directly alters the effective strength of the crab-waist transformation, as described in Eq.~\eqref{eq:k2cw}. The horizontal dispersion modifies the Hamiltonian of $H_{S1}$ to
\begin{equation}
    H_{S1} = \frac{1}{6}K_2 \left[ (x+\eta_x \delta)^3-3(x+\eta_x \delta)y^2 \right],
    \label{eq:Hsext3}
\end{equation}
resulting in tune shift and beta-beat to off-momentum particles. It also perturbs particle motion by modifying Eqs.~\eqref{eq:xmotion}-\eqref{eq:zmotion}, thereby driving synchrobetatron resonances through its interaction with beam-beam effects.

\subsubsection{Imperfect transfer maps between IP and crab-waist sextupoles}

The linear components of $H_{R,L}$ in Fig.~\ref{LieModel} are determined by the optics functions at the IP and the crab-waist sextupole locations. Consequently, the linear imperfections in $H_{R,L}$ correspond to those discussed in Secs.~\ref{sec:ipi} and~\ref{sec:cwsexti}. However, the phase advances between the IP and the crab-waist sextupoles must be considered separately. If these phase advances deviate from the ideal values for the crab-waist transformation, additional terms will arise, modifying Eq.~\eqref{eq:Hcwe1} to
\begin{align}
    H_{cw}^e= &-\frac{1}{2} K_2 \left[ x \sqrt{\frac{\beta_x}{\beta_x^*}} \cos\mu_x + p_x\sqrt{\beta_x\beta_x^*} \sin\mu_x \right] \left[ y \sqrt{\frac{\beta_y}{\beta_y^*}} \cos\mu_y + p_y\sqrt{\beta_y\beta_y^*} \sin\mu_y \right]^2 \nonumber \\
    & +\frac{1}{6} K_2 \left( x \sqrt{\frac{\beta_x}{\beta_x^*}} \cos\mu_x + p_x\sqrt{\beta_x\beta_x^*} \sin\mu_x \right)^3,
    \label{eq:Hcwe2}
\end{align}
where $(\mu_x,\mu_y)$ indicate the phase advances between IP and the crab-waist sextupoles, and the $\alpha$-functions at both locations are assumed to be zero. Setting $(\mu_x,\mu_y)=(\pi,\pi/2)$ simplifies Eq.~\eqref{eq:Hcwe2} to Eq.~\eqref{eq:Hcwe1}.

Nonlinear components can arise in $H_{R,L}$ due to the highly nonlinear IR with extremely small $\beta_y^*$. Nonlinear IR was found to remarkably limit luminosity performance in the final design of SuperKEKB, where $\beta_y^*=0.27$ mm for the LER~\cite{hirosawa2018}.

\subsection{Linking analytic theories with numerical simulations and experimental results of beam-beam effects}

The analytic theories formulated in Sec.~\ref{sec:icwt}, extended to include the machine imperfections outlined in Sec.~\ref{sec:mi}, can be applied to interpret numerical simulations and experimental results of beam-beam effects in crab-waist colliders. This approach is particularly valuable for analyzing weak-strong beam-beam simulations (see, for example, the demonstrations in~\cite{kicsiny2024}). Dedicated simulations with artificially introduced perturbations can be performed to determine the relevant tolerances for machine imperfections, as demonstrated in~\cite{zhou2010ipac}.

In contrast to high-bunch-current collision (HBCC) experiments, which should be compared with strong-strong beam-beam simulations as analyzed in~\cite{zhou2023prab}, weak-strong beam-beam (WSBB) experiments can be conducted to compare with weak-strong simulations. These experiments are particularly useful for identifying potential machine imperfections, especially when using the designed machine tuning knobs. Using SuperKEKB as an example, the WSBB experiment can be designed step by step as follows:
\begin{enumerate}
    \item Perform single-beam tunings to minimize the vertical emittances for both the low-energy ring (LER) and the high-energy ring (HER). This is typically done with multi-bunch mode with low bunch currents.
    \item Refer to Eq.~\eqref{eq:xiy} to estimate the HER bunch current (the strong beam current) for a target beam-beam parameter in the LER beam. For example, achieving $\xi_{y+}=0.05$ in the LER beam requires an HER beam with $I_{b-}=0.63$ mA and $\epsilon_{y-}=30$ pm, assuming $\beta_{y\pm}^*$=0.9 mm. Inject $N_b$ bunches into HER and LER for collision. The LER bunch current should be kept low, for example, around $I_{b+}=$ 0.1 to 0.2 mA (the weak beam current), but high enough to enable accurate measurements of the luminosity and beam sizes.
    \item Perform collision tunings (mainly for the LER beam) and measure luminosity, beam sizes, and other relevant parameters with beam-beam effects in place.
\end{enumerate}

The WSBB experiments focus on the interplay between beam-beam interactions (from the strong beam) and machine imperfections (as discussed in this paper). In HBCC experiments, achieving collisions with balanced beam sizes at the IP is challenging (see~\cite{zhou2023prab}). However, in WSBB experiments, the beams are more controllable: the strong beam experiences negligible beam-beam kick from the weak beam, and the weak beam is either unaffected or only slightly affected by its own single-bunch collective effects (such as impedance, intra-beam scattering, space charge, etc.).

The following items can be quickly investigated through WSBB experiments:
\begin{enumerate}
    \item A survey of LER tune with collisions to compare the results with weak-strong simulations. Any beam blowup observed in the LER beam but not predicted by simulations can be traced to specific sources of machine imperfections.
    \item IP optics knobs to identify IP aberrations. Specifically, closed orbit, waist, linear and chromatic couplings~\cite{zhou2010prstab}, and dispersions at the IP can be adjusted to minimize the blowups in the LER beam.
\end{enumerate}

Advanced machine tunings that can be considered in WSBB experiments, though potentially challenging to implement, include:
\begin{itemize}
    \item Optics knobs at the locations of crab-waist sextupoles: Scan the closed orbit and dispersions at the crab-waist sextupoles to observe their effects on luminosity and beam size blowup.
    \item IP to crab-waist sextupole betatron phase scan: This helps to assess the sensitivity of the crab-waist transformation to betatron phase advances to IP. Note that the global tune should remain fixed during this study.
    \item IR nonlinear knobs: Utilize high-order correctors in the IR to study the effects of IR nonlinearity.
    \item Crab-waist strength ($\chi$) scan: This helps identify the optimal crab-waist strength to maximize luminosity in the presence of nonlinear beam-beam forces. This scan should be performed only after the closed orbit and linear optics at both the IP and the crab-waist sextupoles have been optimized.
\end{itemize}

In fact, all the above-mentioned items can be simulated in advance to assess their relative importance. This approach helps to determine whether implementing these adjustments or conducting the corresponding experiments is necessary for the actual machine.

\subsection{Summary}

The ideal crab-waist transformation effectively mitigates beam-beam-induced betatron resonances, as predicted by beam-beam resonance theory. However, synchrobetatron resonances remain challenging to suppress with the crab-waist scheme. Machine imperfections can disrupt particle motion and beam distribution near the interaction point (IP), introducing additional resonance terms that can degrade the luminosity. A systematic approach, which combines theoretical analysis, simulations, and experimental investigations, allows controlled complexity in the study of these imperfections, ensuring a comprehensive understanding of their impact on performance in crab-waist colliders.

\subsection{ACKNOWLEDGEMENTS}
The author expresses gratitude to K. Ohmi (KEK/IHEP), Z. Li (IHEP), P. Kicsiny (CERN), X. Buffat (CERN), S. Li (USTC), and M. Zobov (INFN) for their insightful discussions on beam-beam effects in crab-waist colliders. Special thanks are extended to M. Zobov for his meticulous proofreading of the manuscript and valuable improvement suggestions. The author also appreciates the SuperKEKB team, as this work greatly benefited from their expertise and hands-on experience with machine observations and tuning in the SuperKEKB control room.



\begin{thebibliography}{999}

    \bibitem{raimondi2006}
        P. Raimondi, 2nd SuperB Workshop, INFN-LNF, Frascati, March, 2006.

    \bibitem{zobov2010test}
        M. Zobov \textit{et al.}, {\em Phys. Rev. Lett} 104, 174801 (2010).

    \bibitem{dikansky2009}
        N.S. Dikansky and D.V. Pestrikov, {Effect of the crab waist and of the micro-beta on the beam–beam instability}, {\em Nuclear Instruments and Methods in Physics Research Section A} 600 (2009) 538-544.

    \bibitem{kicsiny2024}
        P. Kicsiny, D. Zhou, X. Buffat, P. Pieloni, and M. Seidel, ``Incoherent horizontal emittance growth due to the interplay of beam-beam and longitudinal wakefield in crab-waist colliders'', to be published.

    \bibitem{zhou2023prab}
        D. Zhou, K. Ohmi, Y. Funakoshi, Y. Ohnishi, and Y. Zhang, ``Simulations and experimental results of beam-beam effects in SuperKEKB'', {\em Phys. Rev. Accel. Beams} 26, 071001 (2023).

    \bibitem{zhou2024nima}
        D. Zhou, T. Ishibashi, G. Mitsuka, M. Tobiyama, K. Bane, and L. Zhang, ``Theories derived from Haissinski equation and their applications to electron storage rings'', {\em Nuclear Instruments and Methods in Physics Research Section A}, Volume 1063, 169243 (2024).

    \bibitem{handbook2023}
        A. Chao, M. Tigner, H. Weise, and F. Zimmermann, ``Handbook of accelerator physics and engineering'', {\em World scientific}, 2023.

    \bibitem{tawada2005mac}
        M. Tawada, ``Beam-beam update'', presented at the 10th KEKB Accelerator Review Committee meeting, KEK, 2005.

    \bibitem{zhou2024jst}
        D. Zhou, K. Ohmi, Y. Funakoshi, and Y. Ohnishi, ``Luminosity performance of SuperKEKB'', {\em Journal of Instrumentation} 19.02 (2024): T02002.

    \bibitem{ohnishi2022mac}
        Y. Ohnishi, ``Overview of Commissioning in 2021c - 2022b'', presented at the 26th KEKB Accelerator Review Committee meeting, KEK, 2022.

    \bibitem{sugimoto2022mac}
        H. Sugimoto, ``Optics issues'', presented at the 26th KEKB Accelerator Review Committee meeting, KEK, 2022.

    \bibitem{hirosawa2018}
        K. Hirosawa, K. Ohmi, N. Ohuchi, T. Okada, D. Zhou, and N. Kuroo, ``The influence of higher order multipoles of IR magnets on luminosity for SuperKEKB'', In {\em Journal of Physics: Conference Series} (Vol. 1067, No. 6, p. 062004), IOP Publishing (2018).

    \bibitem{zhou2010ipac}
        D. Zhou, K. Ohmi, Y. Seimiya, A. Morita, Y. Ohnishi, and H. Koiso, ``Effects of Linear and Chromatic X-Y Couplings in the SuperKEKB'', in Proc. IPAC'10.

    \bibitem{zhou2010prstab}
        D. Zhou, K. Ohmi, Y. Seimiya, Y. Ohnishi, A. Morita, and H. Koiso, ``Simulations of beam-beam effect in the presence of general chromaticity'', {\em Physical Review Special Topics-Accelerators and Beams} 13, 021001 (2010).

\end{thebibliography}
\end{document}